\documentclass[11pt]{article}
\usepackage{moriond,epsfig}

\def\be{\begin{equation}}
\def\ee{\end{equation}}
\def\bea{\begin{eqnarray}}
\def\eea{\end{eqnarray}}

\begin{document}
\vspace*{4cm}
\title{CHARGED HIGGS IN MODELS WITH SINGLET NEUTRINO\\ 
IN LARGE EXTRA DIMENSIONS \footnote{based on work done in collaboration with 
K. Assamagan. Talk given at XXXVIIth Rencontres de Moriond, ElectroWeak 
Interactions and Unified Theories, March 9th - 16th 2002, Les Arcs, France}
}

\author{ A. DEANDREA}

\address{Institut de Physique Nucl\'eaire, Universit\'e de Lyon I\\  
4 rue E.~Fermi, F-69622 Villeurbanne Cedex, France}

\maketitle
\abstracts{
The charged Higgs decay in the channel $H^-\rightarrow\tau_L^-\nu$ in models 
with a singlet neutrino can provide a test of
large extra dimensions models with TeV scale quantum gravity
since in the standard two Higgs doublet model type II, 
$H^-\rightarrow\tau_L^-\nu$ is suppressed.
In the present study, we focus on the decay $H^-\rightarrow\tau_L^-\psi$ 
at the LHC for Higgs masses larger than the top-quark mass. 
}

\section{Introduction}
Models with extra dimensions postulate the existence of 
$\delta$ additional spatial dimensions of size $R$ where gravity and perhaps 
other fields freely propagate while the SM degrees of freedom are confined to 
(3+1)-dimensional wall (4D) of the higher dimensional space. 
The true scale of gravity, or fundamental Planck scale $M_*$, of the 
($4+\delta$)D space time is related to the reduced 4D Planck scale $M_{Pl}$, 
by $M_{Pl}^2 = R^\delta M_*^{\delta+2}$, where $M_{Pl}=2.4 \times 10^{18}$ 
GeV is related to the usual Planck mass $1.2 \times 10^{19}$ GeV 
$=\sqrt{8\pi} M_{Pl}$.
Since no experimental deviations from Newtonian gravity are 
observed at distances above 0.2 mm \cite{expgra}, the extra dimensions must be 
smaller or equal to the sub-millimeter scale with $M_*$ as low as few TeV 
and $\delta \geq 2$.  
 
The spectrum of many extensions of the SM includes a charged Higgs state. 
$H^-$ decays to the right handed 
$\tau^-$ through the $\tau$ Yukawa coupling:
$H^-\rightarrow \tau_R^-\bar{\nu}$.
The $H^-$ decay to left handed $\tau^-$ is completely suppressed in the 
Minimal Supersymmetric Standard Model (MSSM). 
However, in the scenario of singlet neutrino in large extra dimensions, $H^-$ 
can decay to both right handed and left handed $\tau^-$ depending on the 
parameters $M_*$, $m_D$, $\delta$, $m_{H^\pm}$ and $\tan\beta$: 
$H^- \rightarrow \tau_R^-\bar{\nu} +\tau_L^-\psi$,   
where $\psi$ is a bulk neutrino and $\nu$ is 
dominantly a light neutrino with a small admixture of the Kaluza-Klein modes 
\cite{agashe} of the order $mR/|n|$. The measurement of the polarisation 
asymmetry, can be used to distinguish between the ordinary two Higgs Doublet 
Model of type II (2HDM-II) and the 
scenario of singlet neutrino in large extra dimensions.  
 
\section{$H^\pm$ Production and Decays}
\label{sec:prod}
The charged Higgs decay to right handed $\tau$, 
$H^-\rightarrow\tau_R^-\bar{\nu}$ have been 
extensively studied for the LHC~\cite{6,7}. Here we discuss 
the possibility to observe $H^-\rightarrow\tau_L^-\psi$ at the LHC above
the top-quark mass \cite{Assamagan:2001jw}. 
\begin{table*} [h]
\begin{center} 
\begin{minipage}{0.93\linewidth} 
\caption{\label{tab:table1}The parameters used in the current analysis of the 
signal with the corresponding polarisation asymmetry. In general, $H^-$ would 
decay to $\tau^-_L$ and $\tau^-_R$, $H^-\rightarrow\tau_R^-\bar{\nu} + 
\tau_L^-\psi$, depending on the asymmetry. For the purely decay 
$H^-\rightarrow\tau^-_R\bar{\nu}$ (as in MSSM), the asymmetry is $-1$. The 
signal to be studied is $H^-\rightarrow\tau^-_L\psi$. \hfill ~ }  
\end{minipage}
\vbox{\offinterlineskip 
\halign{&#& \strut\quad#\hfil\quad\cr    
\hline 
& &&$M_*$ (TeV) && $\delta_\nu$, $\delta$ 
&& $m_{H^\pm}$ (GeV) && $\tan\beta$ && $A_{LR}$  
&& $m_\nu$ (eV) &\cr 
\hline  
&Sig.-1 && 2 && 4, 4 &&  219.9 &&  30  && $\sim 1$ && 0.5 $10^{-3}$ \cr   
&Sig.-2 && 20 && 3, 3 && 365.4 && 45 && $\sim 1$ && 0.05 &\cr 
&Sig.-3 && 1 && 5, 6 && 506.2 && 4 && $\sim 1$ && 0.05 &\cr   
&Sig.-4 && 100 && 6, 6 && 250.2 && 35&& $\sim -1$ && 0.005 &\cr 
&Sig.-5 && 10 && 4, 5 && 350.0 && 20 && $\sim -1$ && 0.04 &\cr  
&Sig.-6 && 50 && 5, 5 && 450.0 && 25 && $\sim -1$ && 0.04 &\cr  
\hline
}}   
\end{center}  
\end{table*} 
Table~\ref{tab:table1} shows the parameters selected for
the current analysis. We assume a heavy SUSY spectrum with maximal mixing. 
We consider the $2\rightarrow 2$ production 
process where the charged Higgs is produced with a top-quark, $gb\rightarrow 
tH^\pm$. Further, we require the hadronic decay of the top-quark, 
$t\rightarrow Wb\rightarrow jjb$ and the charged Higgs decay to 
$\tau$-leptons. 

The major backgrounds are the single top production 
$gb\rightarrow Wt$, and $t\bar{t}$ production with one $W^+\rightarrow jj$ and 
the other $W^-\rightarrow\tau_L^-\bar{\nu}$. Depending on the polarisation 
asymmetry, $H^-\rightarrow\tau_R^-\bar{\nu}$ will 
contribute as an additional background. In Table~\ref{tab:table2}, we list the 
rates for the signal and for the backgrounds.  

\begin{table*} [h]
\begin{center}  
\begin{minipage}{0.75\linewidth} 
\caption{\label{tab:table2}
The expected rates ($\sigma\times$ BR), for the signal 
$gb\rightarrow t H^\pm$ with $H^-\rightarrow\tau_R^-\bar{\nu}+\tau_L^-\psi$ 
and $t\rightarrow jjb$, and for the backgrounds: $W t$ and $t\bar{t}$ 
with $W^-\rightarrow\tau_L^-\bar{\nu}$ and $W^+\rightarrow jj$. We assume an 
inclusive $t\bar{t}$ production cross section of 590~pb. In the last columns, 
we 
compare the $H^-\rightarrow\tau_R^-\bar{\nu}$ branching ratios in this model 
to the corresponding MSSM branching ratios. \hfill ~} 
\end{minipage}
\vbox{\offinterlineskip 
\halign{&#& \strut\quad#\hfil\quad\cr 
\hline 
&Process && $\sigma\,\times\,$~BR (pb) && BR && BR(MSSM)&\cr  
\hline  
&Signal-1 && 1.56 && 0.73 &&  0.37  &\cr 
&Signal-2 && 0.15 && 1.0 && 0.15 & \cr  
&Signal-3 && 0.04 && 1.0 && 0.01 & \cr  
\hline 
&$t\bar{t}$ && 84.11 &&  &&   &  \cr  
&$W tb$ ($p_T>30$~GeV) && 47.56 &&  &&   &  \cr 
\hline
}} 
\end{center}  
\end{table*}  

Depending on the parameters $M_*$, 
$m_D$, $\delta$, $m_{H^\pm}$ and $\tan\beta$, the $\tau\nu$ decay of the 
charged Higgs can be enhanced or suppressed compared to the MSSM case.  
\begin{figure} 
\epsfysize=8truecm 
\begin{center} 
\epsffile{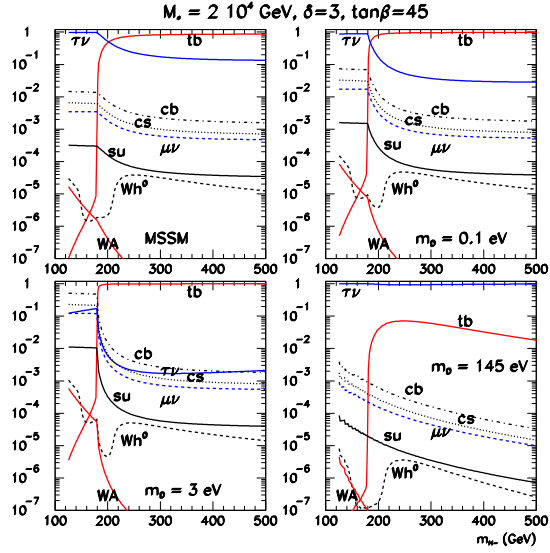} 
\caption{
Charged Higgs decays in models with a singlet neutrino in large extra
dimensions for $M_* = 2\times 10^4$ GeV, $\delta =3$ and $\tan\beta = 45$. 
For small values of $m_D$, we see similar decay branchings as in MSSM. As 
$m_D$ gets larger, $H^\pm\rightarrow\tau\nu$ becomes dominant below and above 
the top-quark mass. \hfill ~}  
\label{fig:led_tb45}  
\end{center} 
\end{figure} 
In Figure~\ref{fig:led_tb45}, we show how the other decays of the charged 
Higgs are affected in this framework; for the chosen values of $M_*$ and 
$\delta$, the decay branchings 
are similar to MSSM for small values of $m_D$ while at larger $m_D$, the 
$\tau\nu$ decay mode becomes strongly enhanced. 

The reconstruction of the transverse mass is not enough to distinguish between 
the MSSM and the singlet neutrinos in large extra dimensions. The differences 
in these two scenarii are best seen in the distribution of 
$p^\pi/E^{\tau-jet}$, the fraction of the energy carried by the charged track. 
\begin{figure} 
\epsfysize=7truecm 
\begin{center} 
\epsffile{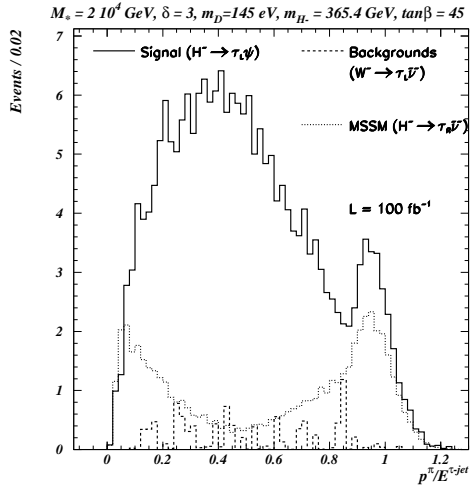} 
\caption{The distribution of the ratio of the charged pion track momentum in 
one prong $\tau$ decay to the $\tau$-jet energy for $m_A=350$~GeV, 
$\tan\beta=45$, $M_*= 20$~TeV, $\delta=3$ and $m_\nu=0.05$~eV. In the 2HDM-II, 
this ratio would peak near 0 and 1 as shown while in other models, the actual 
distribution of this ratio would depend on the polarization asymmetry since 
both left and right handed $\tau$'s would contribute. In the case shown, the 
asymmetry is $\sim 1$ and the ratio peaks near the center of the 
distribution. \hfill ~} \label{fig:350_45_145_1}  
\end{center}  
\end{figure} 

\section{Conclusions} 
 
Large extra dimensions models with TeV scale quantum gravity postulate the 
existence of additional dimensions where gravity (and possibly other fields) 
propagate. The size of the extra dimensions are constrained to the 
sub-millimeter level since no experimental deviations from the Newtonian 
gravity has been observed at distances larger than $\sim 0.2$ millimeter.
The right handed neutrino can freely propagate into the extra 
dimensions because it has no quantum numbers to constrain it to the SM brane. 
The interactions between the bulk neutrino and the 
SM fields on the brane can generate Dirac neutrino masses consistent with the 
atmospheric neutrino oscillations without implementing the seesaw mechanism. 
There are no additional Higgs bosons required in these models. The charged 
Higgs productions are therefore the same as in the two Higgs doublet models.  
The charged Higgs can decay to both the right and the left handed 
$\tau$-leptons, $H^-\rightarrow\tau_R^-\bar{\nu}+\tau_L^-\psi$ whereas in the 
2HDM-II such as MSSM, only the right handed $\tau$ decay of the $H^-$ is 
possible through the $\tau$ Yukawa coupling: 
$H^-\rightarrow\tau_R^-\bar{\nu}$. The $\tau$ decay of 
the charged Higgs has been studied in details for ATLAS and CMS. In the 
current study, we focus on the decay $H^-\rightarrow\tau_L^-\psi$ 
at the LHC for Higgs masses larger than the top-quark mass. 

Although the observation of a signal in the transverse mass 
distribution can be used to claim discovery of the charged Higgs, it is 
insufficient to select the scenario that is realized. Additionally, by 
reconstructing the fraction of the energy carried  by the charged track in the 
one-prong $\tau$ decay, it is possible to distinguish whether the scenario is 
the ordinary two Higgs doublet model or not. The further measurement of the 
polarisation asymmetry might provide a distinctive evidence for models with 
singlet neutrino in large extra dimensions. 

\section*{Acknowledgements} 
I would like to thank K.~A.~Assamagan and Y.~Coadou for fruitful 
collaborations.

\end{document}